\documentclass[aps,preprint]{revtex4-1}

\usepackage{latexsym}
\usepackage{float}

\usepackage{mathrsfs}
\usepackage{graphicx,amsfonts,amssymb}
\usepackage[fleqn]{amsmath}
\usepackage{epstopdf}
\usepackage{hyperref}
\begin{document}


%
%

\title{CRITICAL TEMPERATURE OF PAIR CONDENSATION IN A DILUTE BOSE GAS WITH SPIN-ORBIT COUPLING}

\author{DEKUN LUO, LAN YIN}
\address{School of Physics, Peking University, Beijing 100871, China,
\\
yinlan@pku.edu.cn}


\begin{abstract}
We study the Bardeen-Cooper-Shrieffer (BCS) pairing state of a two-component Bose gas with a symmetric spin-orbit coupling.  In the dilute limit at low temperatures, this system is essentially a dilute gas of diatomic molecules.  We compute the effective mass of the molecule and find that it is anisotropic in momentum space.  The critical temperature of the pairing state is about eight times smaller than the Bose-Einstein condensation (BEC) transition temperature of an ideal Bose gas with the same density.
\end{abstract}


\maketitle

\section{Introduction}
In contrast of the successful observation of BCS-BEC crossover in Fermi gases ,\cite{010101} the BCS pairing state of Bosons has been proposed more than half a century ago ,\cite{ISI:A1958WQ70400002,ISI:A1969E277200015} but it is still not observed in experiments. In experiments of ultracold quantum gases, one big obstacle is the short lifetime of Feshbach molecules ,\cite{010401,010402,010403,010404} not enough for equilibration into a molecular BEC state. Theoretically, the BCS state of a Bose gas was found to be generally unstable in the attractive region or close to the Feshbach resonance.\cite{intro1,intro2,intro3,intro4}  Recently we proposed to realize a stable BCS pairing state of a Bose gas by spin-orbit coupling (SOC) .\cite{2016arXiv160909691L}  In this paper we study the critical temperature of this pairing state.

SOC in ultracold quantum gas has attracted a lot of research attention in recent years. SOC was first realized in Bose gases ,\cite{intro20102,intro20101} and then in Fermi gases .\cite{intro20201,intro20202}  The SOC of ultracold atoms is not SOC of electrons, but the artificial coupling between spin of the atomic internal state and center-of-mass momentum of the atom . \cite{intro20304,intro20305,intro20307,intro20310,intro20304j02}  Although most studies focused on one-dimensional(1D) SOC, two-dimensional(2D) SOC has been recently realized in experiments, \cite{020101,020102} and the methods to generate three-dimensional(3D) SOC have been proposed. \cite{020201,020202}

This paper is organized as follows: First, we introduce the BCS pairing state of a Bose gas with symmetric SOC.  Then we study the two-body bound state of Bose atoms with 3D SOC at finite center-of-mass momentum and calculate the effective mass of this diatomic molecule, which is used in the computation of the critical temperature.  Finally, we discuss experimental aspects and conclude in the end.

\section{Pairing state of a Bose gas with SOC}

\subsection{Introduction}

We consider a two-component homogeneous Bose gas modeled by the Hamiltonian $H=H_{0}+H_{int}$. The single-particle Hamiltonian $H_0$ is given by
\begin{eqnarray}
H_0=\sum_{\textbf{k},\rho,\rho'}c^{\dag}_{\textbf{k}\rho}[\epsilon_{\textbf{k}}  \delta_{\rho\rho'}+\frac{\hbar^{2}\kappa}{m} \textbf{k}\cdot {\bf \sigma}_{\rho\rho'}] c_{\textbf{k}\rho'},
\end{eqnarray}
where ${\bf \sigma}_{\rho\rho'}$ are Pauli matrices, $m$ is the atomic mass, $c_{\textbf{k}\rho}$ is the annihilation operator of a Boson with wavevector $\textbf{k}$ and spin component $\rho$, $\epsilon_k=\hbar^2k^2/2m$, and $\kappa$ is the strength of an isotropic 3D SOC.  The single-particle Hamiltonian $H_0$ can be diagonalized, yielding two helicity excitation branches with eigenenergies given by $\epsilon_k\pm\hbar^{2}\kappa k/m$.  The interaction between atoms $H_{int}$ is given by
\begin{eqnarray}
H_{int}=\frac{1}{2V}\sum_{\textbf{kk}'\textbf{q}\rho\rho'}g_{\rho\rho'}c^{\dag}_{\frac{\textbf{q}}{2}+\textbf{k}'\rho}
c^{\dag}_{\frac{\textbf{q}}{2}-\textbf{k}'\rho'}c_{\frac{\textbf{q}}{2}-\textbf{k}\rho'}c_{\frac{\textbf{q}}{2}+\textbf{k}\rho},
\end{eqnarray}
where V is the volume, $g_{\rho\rho'}$ are coupling constants, and $g_{\uparrow\downarrow}=g_{\downarrow\uparrow}$.  In this paper we consider the symmetric case, $g_{\uparrow\uparrow}=g_{\downarrow\downarrow}$.

With 3D SOC, two Bose atoms can form a bound state with any inter- or intra-species coupling constants \cite{2016arXiv160909691L}, which is helpful for the formation of a BCS paring state.  Phase separation can be avoided with intra-species repulsion and inter-species attraction.  In this case since the binding energy of the inter-species diatomic molecule is much smaller, and it will be easier to generate inter-species diatomic molecules.  At low temperatures, these diatomic molecules condense into a pairing state \cite{2016arXiv160909691L}.  In addition to single-particle excitations described in the mean-field approximation \cite{2016arXiv160909691L}, there are also pair excitations.  In the dilute limit, the pairing state is essentially a weakly interacting BEC state of diatomic molecules at low temperatures \cite{010101,intro301}.

\subsection{Properties of a diatomic molecule with SOC}

The diatomic wavefunction $|\phi\rangle_\textbf{q}$ of two Bose atoms satisfies the eigenequation $H|\phi\rangle_\textbf{q}=E_{\textbf{q}}|\phi\rangle_\textbf{q}$, where $\hbar\textbf{q}$ is the center of mass momentum, and $E_{\textbf{q}}$ is the eigenenergy.  This wavefunction can be generally written as
\begin{eqnarray}
|\phi\rangle_\textbf{q}=\sum_{\textbf{k}\rho\rho'}\psi_{\rho\rho'}(\textbf{k},\textbf{q}-\textbf{k})c^{\dag}_{\textbf{k}\rho}
c^{\dag}_{\textbf{q}-\textbf{k}\rho'}|0\rangle,
\end{eqnarray}
where the coefficients satisfy $\psi_{\rho\rho'}(\textbf{k},\textbf{k}')=\psi_{\rho'\rho}(\textbf{k}',\textbf{k})$.
From the eigenequation, we can construct the following matrix equation for the coefficients
\begin{eqnarray}
M_{\textbf{kq}}\psi'_{\textbf{kq}}=\frac{1}{V}G\sum_{\textbf{p}}\psi'_{\textbf{pq}},
\end{eqnarray}
where $\psi'_{\textbf{kq}}=[\psi_{\uparrow\uparrow}(\frac{\textbf{q}}{2}+\textbf{k}, \frac{\textbf{q}}{2}-\textbf{k}),\psi_{\downarrow\downarrow}(\frac{\textbf{q}}{2}+\textbf{k},\frac{\textbf{q}}{2}-\textbf{k}), \psi_{\uparrow\downarrow}(\frac{\textbf{q}}{2}+\textbf{k},\frac{\textbf{q}}{2}-\textbf{k}),\psi_{\uparrow\downarrow}(\frac{\textbf{q}}{2}+\textbf{k},\frac{\textbf{q}}{2}-\textbf{k})]$,
\begin{eqnarray}
G=\left(
\begin{matrix}
g_{\uparrow\uparrow}&0&0&0\\
0&g_{\downarrow\downarrow}&0&0\\
0&0&g_{\uparrow\downarrow}&0\\
0&0&0&g_{\uparrow\downarrow}
\end{matrix}
\right).
\end{eqnarray}
The matrix $M_{\textbf{kq}}$ is given by
\begin{eqnarray}
M_{\textbf{kq}}=\left(
\begin{matrix}
\varepsilon_{\textbf{kq}}-\frac{\hbar^{2}\kappa q_{z}}{m}&0&-\frac{1}{2}S^{*}(\textbf{q}_{\perp})+S^{*}(\textbf{k}_{\perp})&-\frac{1}{2}S^{*}(\textbf{q}_{\perp})-S^{*}(\textbf{k}_{\perp})\\
0&\varepsilon_{\textbf{kq}}+\frac{\hbar^{2}\kappa q_{z}}{m}&-\frac{1}{2}S(\textbf{q}_{\perp})-S(\textbf{k}_{\perp})&-\frac{1}{2}S(\textbf{q}_{\perp})+S(\textbf{k}_{\perp})
\\
-\frac{1}{2}S(\textbf{q}_{\perp})+S(\textbf{k}_{\perp})&-\frac{1}{2}S^{*}(\textbf{q}_{\perp})-S^{*}(\textbf{k}_{\perp})
&\varepsilon_{\textbf{kq}}-\frac{2\hbar^{2}\kappa k_{z}}{m}&0
\\
-\frac{1}{2}S(\textbf{q}_{\perp})-S(\textbf{k}_{\perp})&-\frac{1}{2}S^{*}(\textbf{q}_{\perp})+S^{*}(\textbf{k}_{\perp})
&0&\varepsilon_{\textbf{kq}}+\frac{2\hbar^{2}\kappa k_{z}}{m}
\end{matrix}
\right),\nonumber\\
\end{eqnarray}
where $\varepsilon_{\textbf{kq}}=E_{\textbf{q}}-(\frac{\hbar^{2}q^{2}}{4m}+\epsilon_k$, $\textbf{k}_{\perp}$ is the projection of $\textbf{k}$ in the $x-y$ plane, and $S(\textbf{k}_{\perp})=\hbar^{2}\kappa(k_{x}+ik_{y})/m$.  Define $Q=G\sum_{\textbf{k}}\psi'_{\textbf{kq}}/V$,
and we obtain
\begin{eqnarray}
Q=\frac{1}{V}G\sum_{\textbf{k}}M_{\textbf{kq}}^{-1}Q.
\end{eqnarray}
Thus the eigenenergy of the bound state satisfies the equation
\begin{eqnarray}
\parallel1-\frac{1}{V}G\sum_{\textbf{k}}M_{\textbf{kq}}^{-1}\parallel=0, \label{bound1}
\end{eqnarray}
which has three nontrivial bound-state solutions, two due to the intra-species interaction and one due to inter-species interaction.

As stated in the previous subsection, we consider only the diatomic molecule due to inter-species interaction.  By solving Eq. (\ref{bound1}), we obtain in the limit $q \rightarrow 0$,
\begin{eqnarray}
E_{\textbf{q}}=E_{0}+\frac{\hbar^{2}q_z^{2}}{2m_z^{*}}+\frac{\hbar^{2}q_\perp^{2}}{2m_\perp^{*}},
\end{eqnarray}
where $m_z^{*}$ and $m_\perp^{*}$ are axial and planar effective masses.  From numerically computation, we find that the effective mass is anisotropic, $m_z^{*} \neq m_\perp^{*}$.  The planar effective mass reaches maximum when $a_{\uparrow\downarrow}$ diverges, as shown in Fig.1.  We find in the limit $a_{\uparrow\downarrow}\rightarrow 0^{-}$, $m_\perp^{*}=10m$ and $m_z^{*}/m=10/3$ ; at resonance, $m_\perp^{*}=16.5m$ and $m_z^{*}=2.5m$; in the limit $a_{\uparrow\downarrow}\rightarrow 0^{+}$, $m^{*}_\perp=m^{*}_z=2m$. For $\kappa a_{\uparrow\downarrow}=-1.49$ good for experiments \cite{2016arXiv160909691L}, we obtain
$m_\perp^{*}=14.2m$ and $m_z^{*}=2.9m$.

\begin{figure}
\includegraphics[width=12cm]{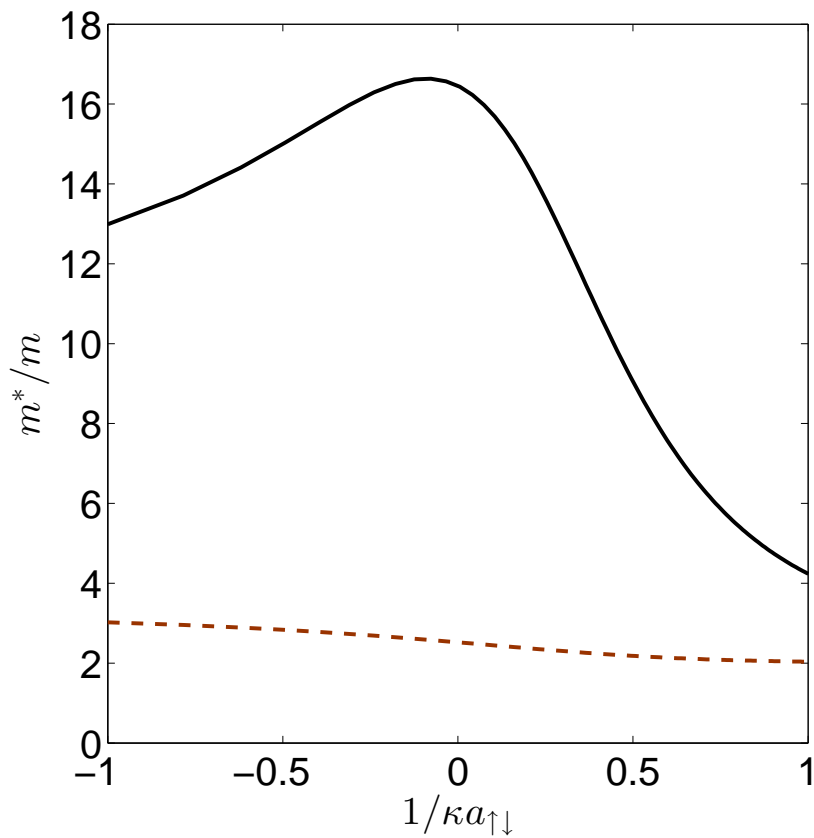}
\caption{Effective masses of a diatomic molecule vs inverse of inter-species scattering length.  The solid and dashed lines are planar and axial effective masses respectively.}
\end{figure}

\subsection{Critical temperature of the pairing state}
In the dilute limit, the interaction between diatomic molecules is very weak.  At the critical temperature $T_{c}$, almost all the atoms form thermal molecules with density
\begin{equation}\label{Tc1}
n=\frac{1}{V}\sum_{\textbf{q}}\ \frac{1}{e^{\beta(E_{\textbf{q}}-E_{0})}-1},
\end{equation}
where $\beta=\frac{1}{k_{B}T_{c}}$.  In the effective mass approximation, we obtain
\begin{equation}
T_{c} \approx \frac{m}{(m_z^{*}{m_\perp^{*}}^2)^{1/3}}T_{a},
\end{equation}
where $T_{a}=2\pi \zeta^{-\frac{2}{3}}(\frac{3}{2})\hbar^{2}(n)^{\frac{2}{3}}k_{B}^{-1}m^{-1}$ is the critical temperature of an ideal two-component Bose gas with density $2n$, and $\zeta$ is the Riemann-zeta function.

For $\kappa a_{\uparrow\downarrow}=-1.49$, we obtain that $T_{c}=0.12T_{a}$, about eight times smaller than the BEC transition temperature of an ideal Bose gas.  For an ultra-cold $^{87}$Rb gas with $2n=2\times10^{17}m^{-3}$, $\kappa a_{\uparrow\downarrow}=-1.49$, and $\kappa=2.51\times10^{7}m^{-1}$, the critical temperature of the BCS pairing state can be observed at about $0.76nK$ which is already reachable with current experimental techniques.

\section{Summary}

In summary, we study the critical temperature of the BCS pairing state in a two-component Bose gas with a symmetric spin-orbit coupling.  In the dilute limit at low temperatures, this system is a weakly interacting gas of diatomic molecules.  We find that the effective mass of a diatomic molecules is anisotropic.  The critical temperature of the pairing state is about eight times smaller than the BEC transition temperature of an ideal Bose gas with the same density.

\section*{Acknowledgements}

We would like to thank T.-L. Ho, Z.-Q. Yu and Rong Li for helpful discussions.  This work is supported by NSFC under Grant No. 11274022 and NKRDP under Grant No. 2016YFA0301500.

\end{document}